\documentclass[a4paper,final]{appolb}
\pdfoutput=1
\usepackage[pdftex]{graphicx}
\usepackage[unicode=true, bookmarks=false,bookmarksnumbered=true, bookmarksopen=true, breaklinks=false, backref=false, pagebackref=false, colorlinks=true]{hyperref}
\hypersetup{pdftitle={Flux tubes at Finite Temperature}, pdfauthor={Nuno Cardoso}, linkcolor=black, citecolor=black, urlcolor=black, filecolor=black, pdfpagelayout=OneColumn, pdfnewwindow=true, pdfstartview=XYZ, plainpages=false}
\usepackage{color}
\usepackage{eurosym}
\usepackage{bm}
\usepackage{amsmath}
\usepackage{amssymb}
\usepackage{amsmath, amsfonts, epsfig, xspace}
\usepackage{array}
\usepackage{tabularx}
\usepackage{booktabs}
\usepackage{enumerate}
\usepackage{algorithm,algorithmic}
\usepackage{pgf}
\usepackage{verbatim}
\usepackage{braket}
\usepackage[T1]{fontenc}
\usepackage[english]{babel}
\usepackage{subfig}
\usepackage{float}
\usepackage{url}
\usepackage{listings}
\usepackage{algorithm}
\usepackage{algorithmic}
\usepackage{grffile}

\usepackage{adjustbox}
\usepackage{array}
\usepackage{booktabs}
\usepackage{multirow}

\newcommand\T{\rule{0pt}{2.0ex}}
\newcommand\B{\rule[-0.6ex]{0pt}{0pt}}

\graphicspath{{figures/}}

\usepackage{verbatim}

\usepackage{cleveref}
\crefname{section}{Section}{Sections}
\Crefname{section}{Section}{Sections}
\crefname{equation}{Eq.}{Eqs.}
\Crefname{equation}{Eq.}{Eqs.}
\crefname{figure}{Fig.}{Figs.}
\Crefname{figure}{Fig.}{Figs.}
\crefname{table}{Table}{Tables}
\Crefname{table}{Table}{Tables}

\usepackage{graphicx}
\begin{document}

\title{Flux tubes at Finite Temperature
\thanks{Presented by N. Cardoso at the International Meeting "Excited QCD", Costa da Caparica, Portugal, 6 - 12 March, 2012}}
\author{Nuno Cardoso, Marco Cardoso and Pedro Bicudo
\address{CFTP, Departamento de F\'{i}sica, Instituto Superior T\'{e}cnico, Universidade de Lisboa, Av. Rovisco Pais, 1049-001 Lisbon, Portugal}
}

\maketitle

\begin{abstract}
In this work, we show the flux tubes of the quark-antiquark and quark-quark at finite temperature for SU(3) Lattice QCD.
The chromomagnetic and chromoelectric fields are calculated above and below the phase transition.
\end{abstract}
\PACS{11.15.Ha; 12.38.Gc}

\section{Introduction}

The study of the chromo fields distributions inside the flux tubes formed $QQ$ and $Q\bar{Q}$ are presented in this study.
How the flux tube evolves when the distance between quarks or the temperature increase beyond respective
critical values are addressed in this paper.
In section 2, we describe the lattice formulation. We briefly review the Polyakov loop for these systems and show how to compute the color fields as well as the Lagrangian distribution. In section 3, the numerical results are shown. Finally, we conclude in section 4.

\section{Computation of the chromo-fields in the flux tube}

The central observables that govern the event in the flux tube can be extracted from the correlation of a plaquette, $\square_{\mu\nu}$, with the Polyakov loops, $L$,
\begin{equation}
	f_{\mu\nu}(r,x) = \frac{\beta}{a^4} \left[\frac{\Braket{\mathcal{O}\,\square_{\mu\nu}(x)}}{\Braket{\mathcal{O}}}-\Braket{\square_{\mu\nu}(x)}\right]
\end{equation}
where $\mathcal{O}=L(0)\,L^\dagger(r)$ for the $Q\bar{Q}$ system or $\mathcal{O}=L(0)\,L(r)$ for the $QQ$ system,
$x$ denotes the distance of the plaquette from the line connecting quark sources, $r$ is the quark separation, $L(r)=\frac{1}{3}\Tr\Pi_{t=1}^{N_t}U_4(r,t)$
where $N_t$ is the number of time slices of the lattice and using the periodicity in the time direction for the plaquette, 
$\square_{\mu\nu}(x) = \frac{1}{N_t} \sum_{t=1}^{N_t} \square_{\mu\nu}(x,t)$, allows averaging over the time direction.

To reduce the fluctuations of the $\mathcal{O}\,\square_{\mu\nu}(x)$, we measure the following quantity, \cite{PhysRevD.47.5104},
\begin{equation}
	f_{\mu\nu}(r,x) = \frac{\beta}{a^4} \left[\frac{\Braket{\mathcal{O}\,\square_{\mu\nu}(x)}-\Braket{\mathcal{O}\,\square_{\mu\nu}(x_R)}}{\Braket{\mathcal{O}}}\right]
	\label{eq:fmunucomp}
\end{equation}
where $x_R$ is the reference point placed far from the quark sources.

Therefore, using the plaquette orientation $(\mu,\nu)=(2,3), (1,3), (1,2),$ $(1,4),(2,4),(3,4)$, we can relate the six components in \cref{eq:fmunucomp} to the components of the chromoelectric and chromomagnetic fields,
\begin{equation}
	f_{\mu\nu}\rightarrow\frac{1}{2}\left(-\Braket{B_x^2},-\Braket{B_y^2},-\Braket{B_z^2},\Braket{E_x^2},\Braket{E_y^2},\Braket{E_z^2}\right)
\end{equation}
and also calculate the total action (Lagrangian) density, $\Braket{\mathcal{L}}=\frac12\left(\Braket{E^2}-\Braket{B^2}\right)$

In order to improve the signal over noise ratio, we use the multihit technique,  \cite{Brower:1981vt, Parisi:1983hm}, replacing each temporal link by it's thermal average,
and the extended multihit technique, \cite{Cardoso:2013lla}, which consists in replacing  each temporal link by it's thermal average with the first $N$ neighbors fixed. 
Instead of taking the thermal average of a temporal link with the first neighbors, we fix the higher order neighbors, and apply the heat-bath algorithm to all the links inside, averaging the central link,
\begin{equation}
	U_4\rightarrow \bar{U}_4=\frac{\int \left[\mathcal{D}U_4\right]_\Omega U_4 \,e^{\beta \sum_{\mu,s} \Tr\left[ U_\mu(s) F^\dagger(s)\right]}}{\int \left[\mathcal{D}U_4\right]_\Omega \,e^{\beta \sum_{\mu,s} \Tr\left[ U_\mu(s) F^\dagger(s)\right]}}
\end{equation}
By using $N = 2$ we are able to greatly improve the signal, when compared with the error reduction achieved with the simple multihit. Of course, this technique is more computer intensive than simple multihit, while being
simpler to implement than multilevel.
The only restriction is $R > 2N$ for this technique to be valid.

\section{Results}

In this section, we present the results for different $\beta$ values suing a fixed lattice volume of $48^3\times 8$, \cref{tab:latticesimdata}.
All the computations were done in NVIDIA GPUs using CUDA.

\begin{table}[!htb]
\begin{center}
\begin{tabular}{|cccc|}
\hline
\T\B $\beta$ & $T/T_c$	&	$a\sqrt{\sigma}$	&	\# config.	\\ \hline
\hline
\T\B 5.96 & 0.845	&	0.235023	&	5990	\\ \hline
\T\B 6.055 & 0.988	&	0.200931	&	5990/4775* 	\\ \hline
\T\B 6.1237 & 1.100	&	0.180504	&	3669 	\\ \hline
\T\B 6.2 & 1.233	&	0.161013	&	1868 	\\ \hline
\T\B 6.338 & 1.501	&	0.132287	&	3688 	\\ \hline
\T\B 6.5 & 1.868	&	0.106364	&	1868 	\\ \hline
\end{tabular}
\end{center}
\caption{Lattice simulations for a $48^3\times 8$ volume. The lattice spacing was computed using the parametrization from \cite{Edwards:1997xf} in units of the string tension at zero temperature. The $^*$ means without configurations in the wrong phase transition.}
\label{tab:latticesimdata}
\end{table}

\begin{figure}[htp]
	\captionsetup[subfloat]{farskip=0.5pt,captionskip=0.5pt}
\begin{centering}
\subfloat[$\beta=5.96$, $T=0.845T_c$.\label{fig:F00_5.96_ppdagger_}]{
\begin{centering}
\includegraphics[trim=56 30 10 0, clip,width=6.2cm]{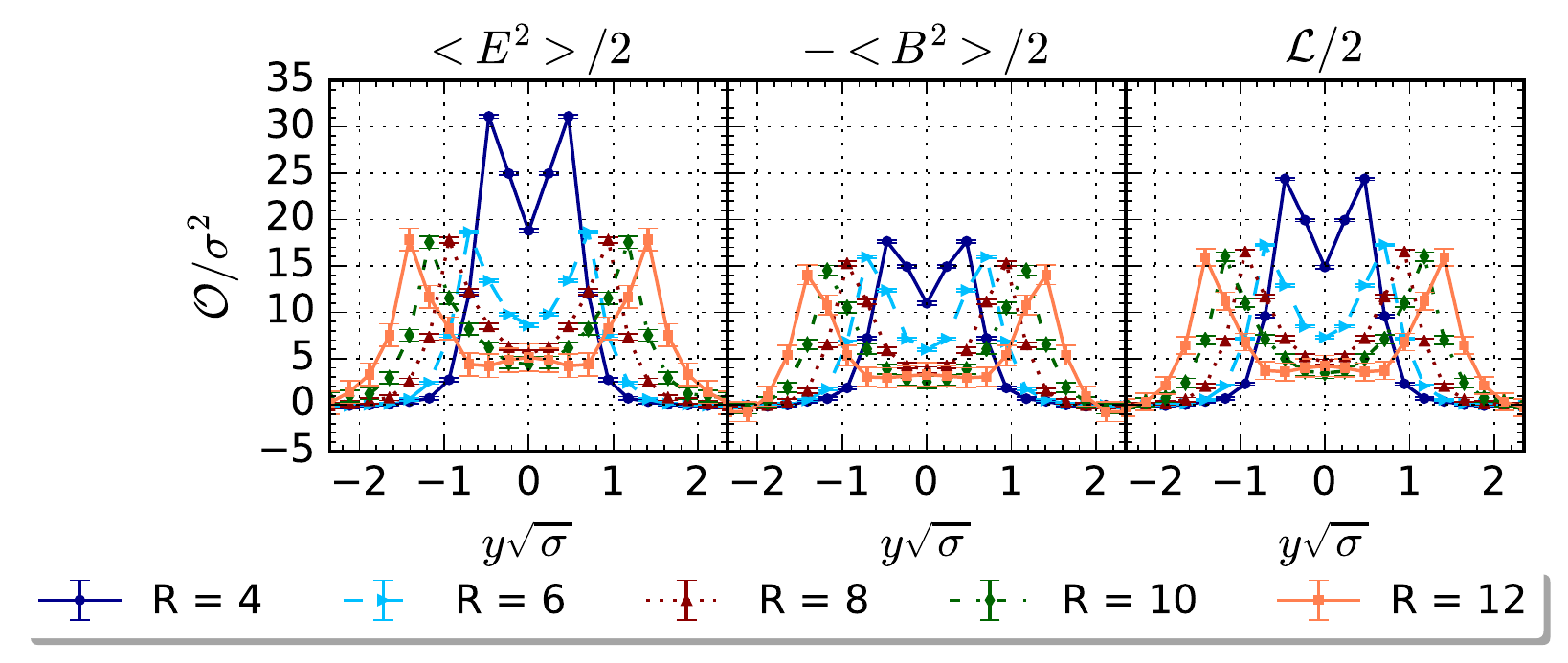}
\includegraphics[trim=56 30 10 0, clip,width=6.2cm]{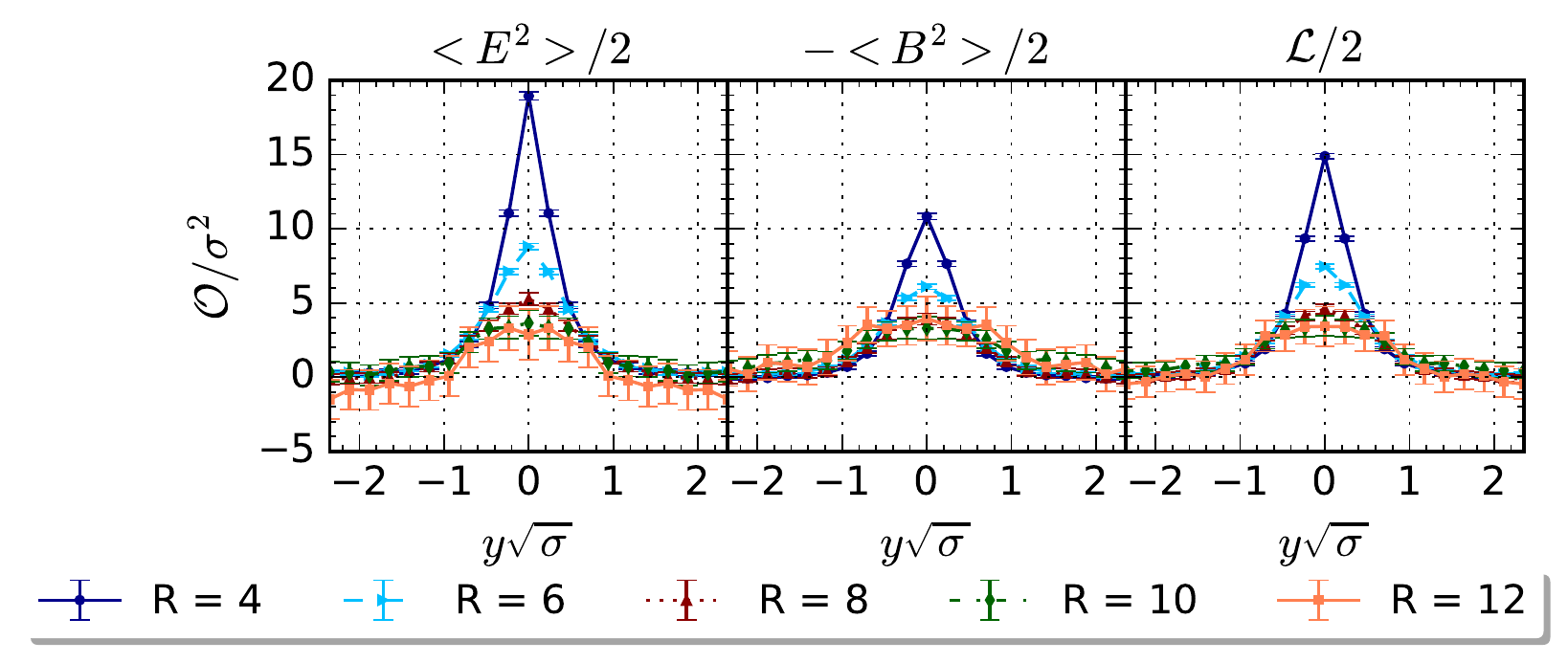}
\par\end{centering}}

\subfloat[$\beta=6.055$, $T=0.988T_c$, with contaminated configurations.\label{fig:F00_6.055_ppdagger_}]{
\begin{centering}
\includegraphics[trim=56 30 10 0, clip,width=6.2cm]{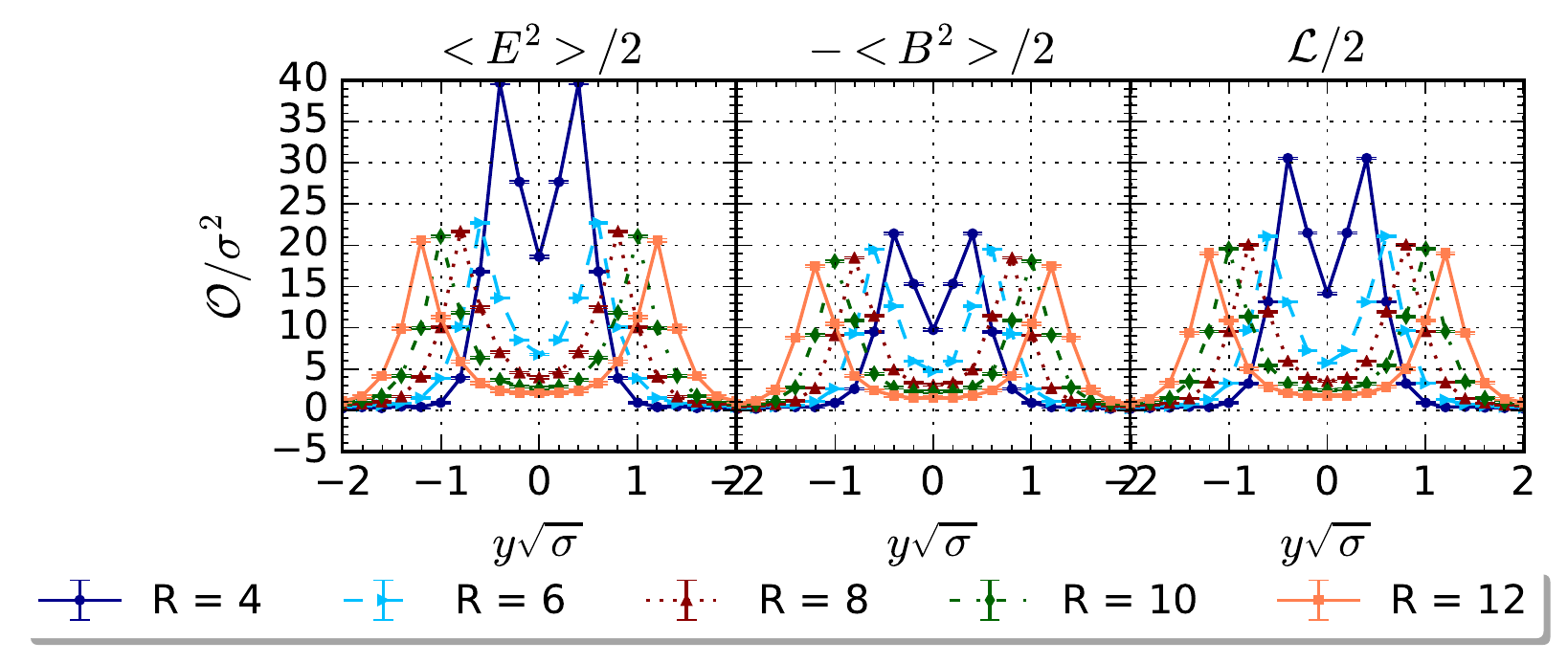}
\includegraphics[trim=56 30 10 0, clip,width=6.2cm]{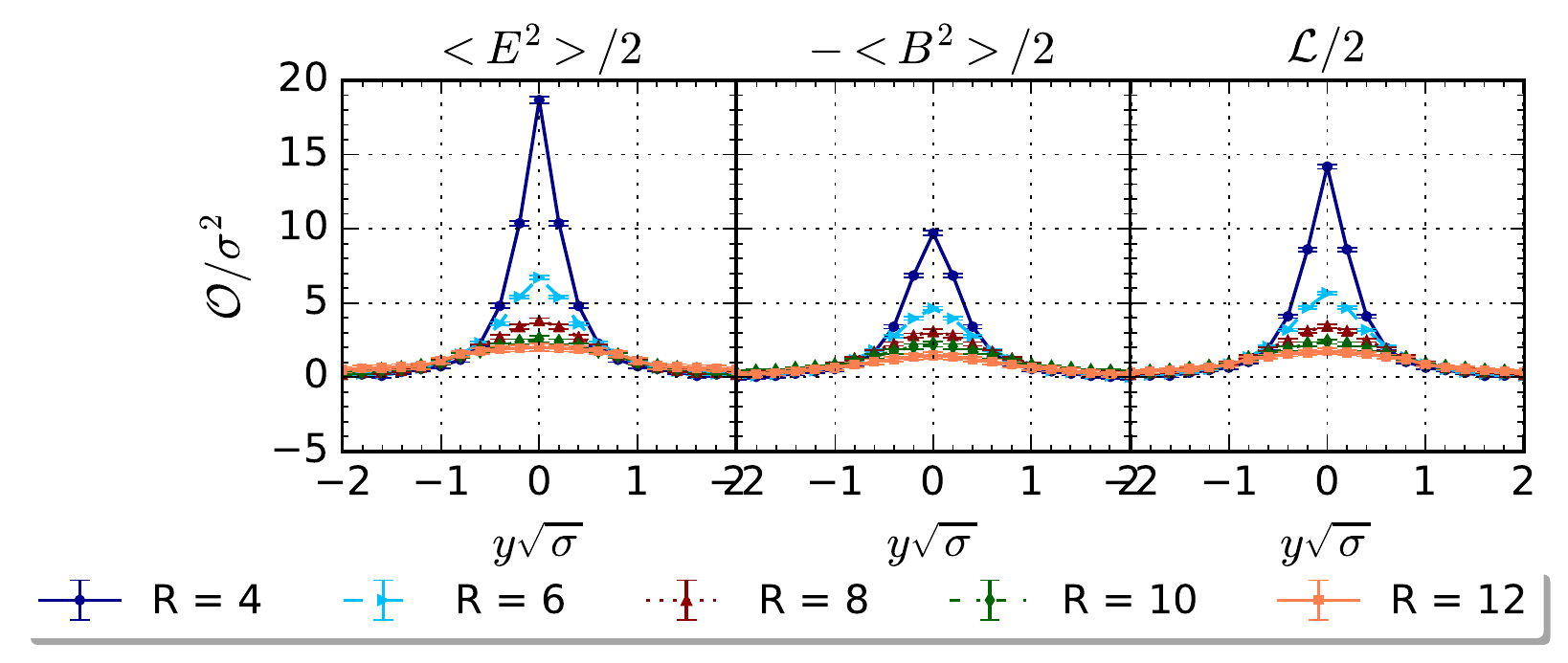}
\par\end{centering}}

\subfloat[$\beta=6.055$, $T=0.988T_c$, without contaminated configurations.\label{fig:F00_clean_6.055_ppdagger_}]{
\begin{centering}
\includegraphics[trim=56 30 10 0, clip,width=6.2cm]{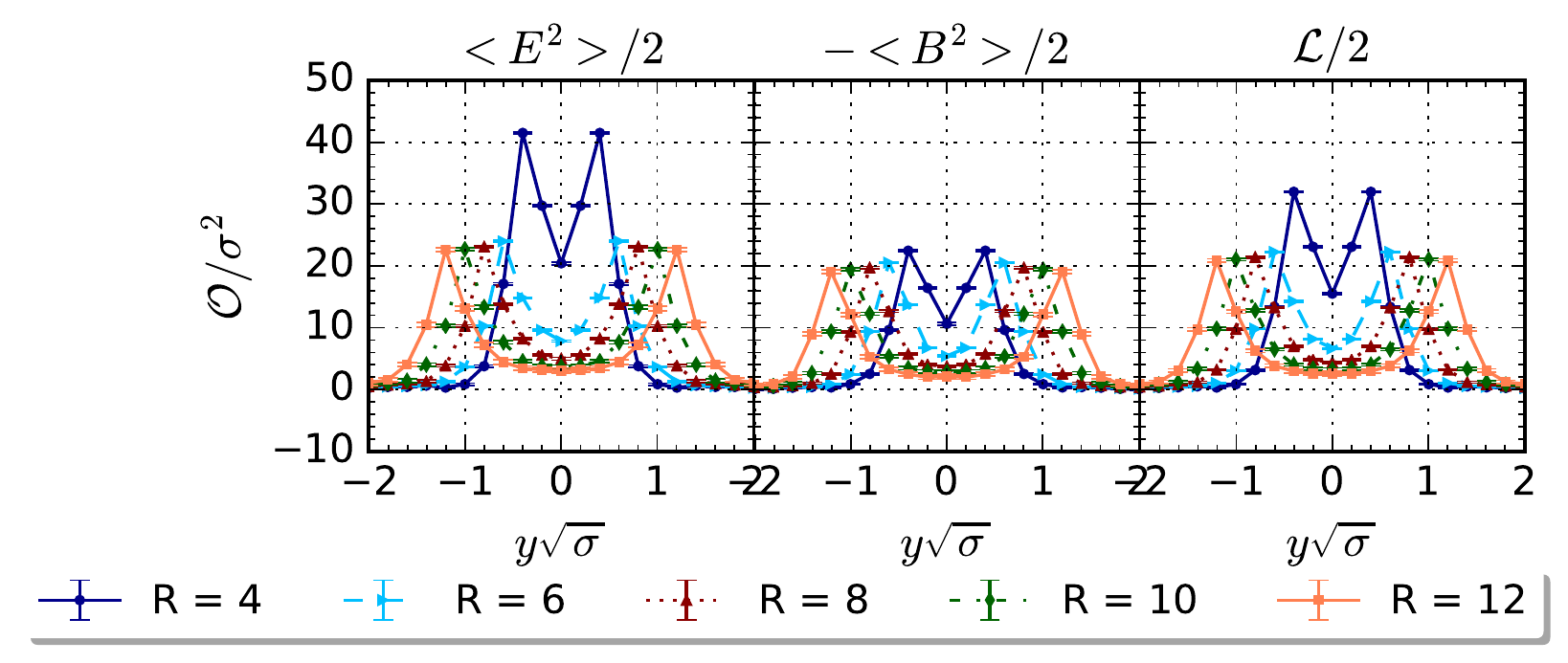}
\includegraphics[trim=56 30 10 0, clip,width=6.2cm]{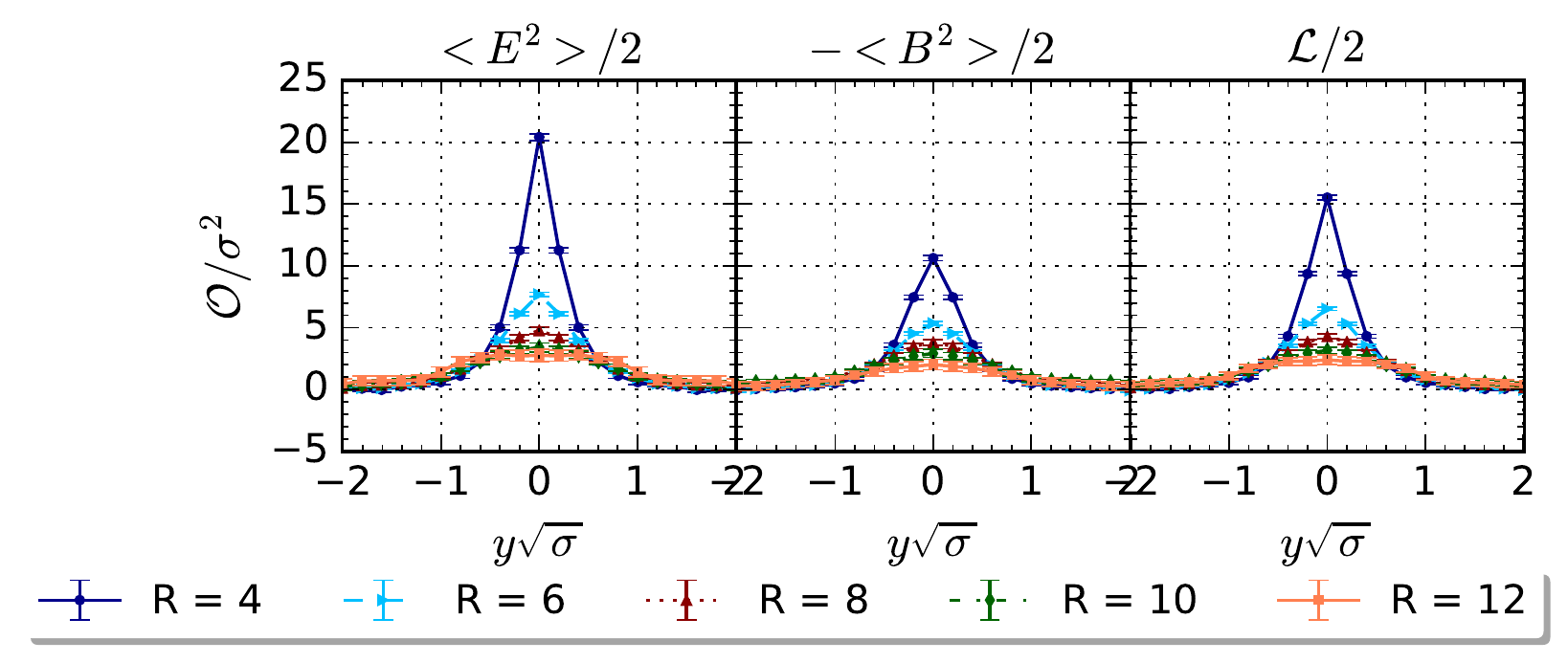}
\par\end{centering}}

\subfloat[$\beta=6.2$, $T=1.233T_c$.\label{fig:F00_6.2_ppdagger_}]{
\begin{centering}
\includegraphics[trim=56 30 10 0, clip,width=6.2cm]{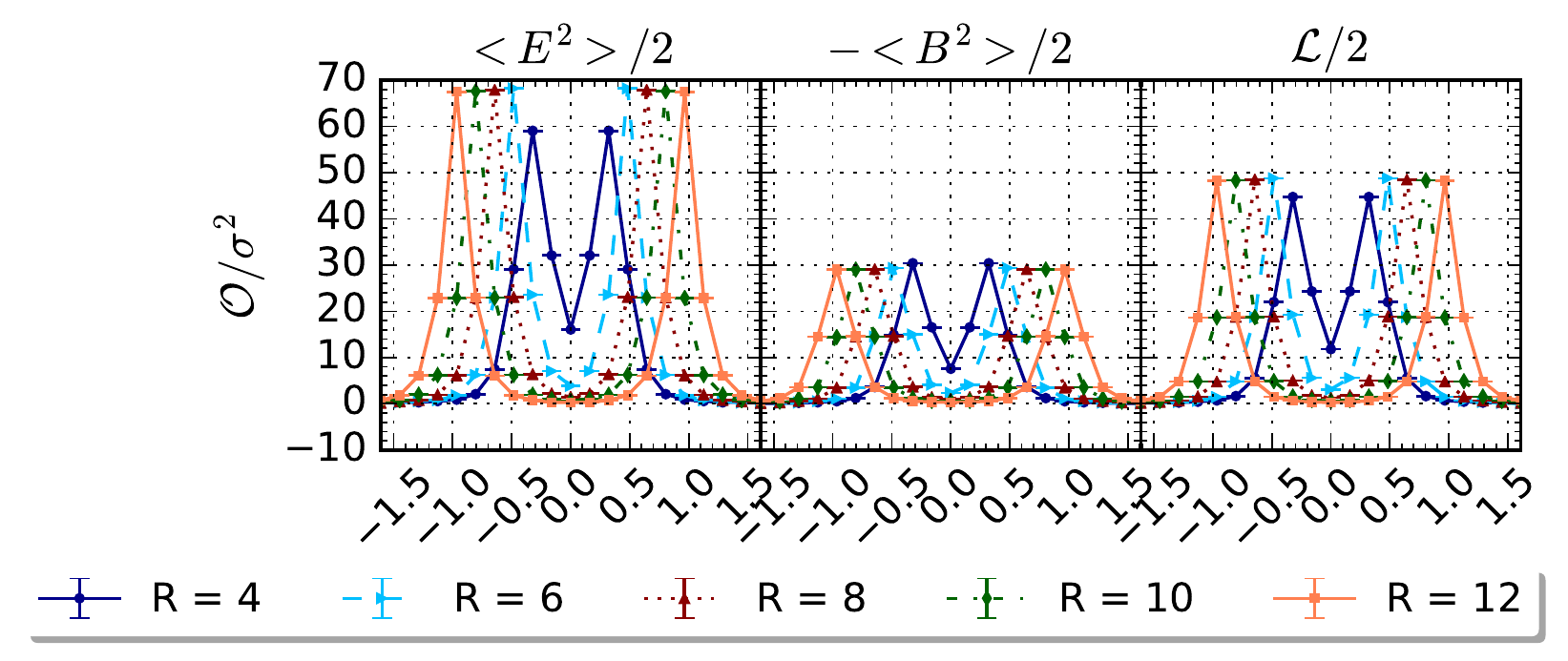}
\includegraphics[trim=56 30 10 0, clip,width=6.2cm]{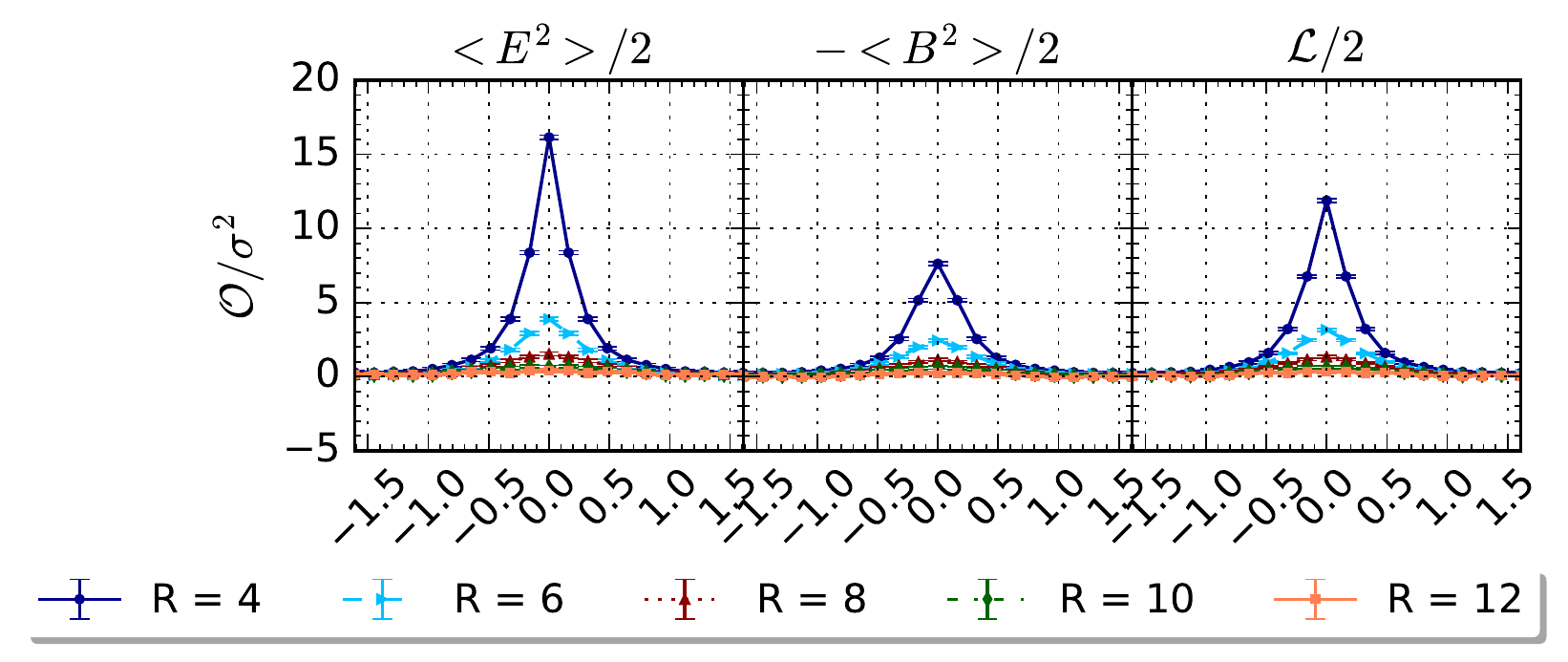}
\par\end{centering}}

\subfloat[$\beta=6.5$, $T=1.868T_c$.\label{fig:F00_6.5_ppdagger_}]{
\begin{centering}
\includegraphics[trim=56 30 10 0, clip,width=6.2cm]{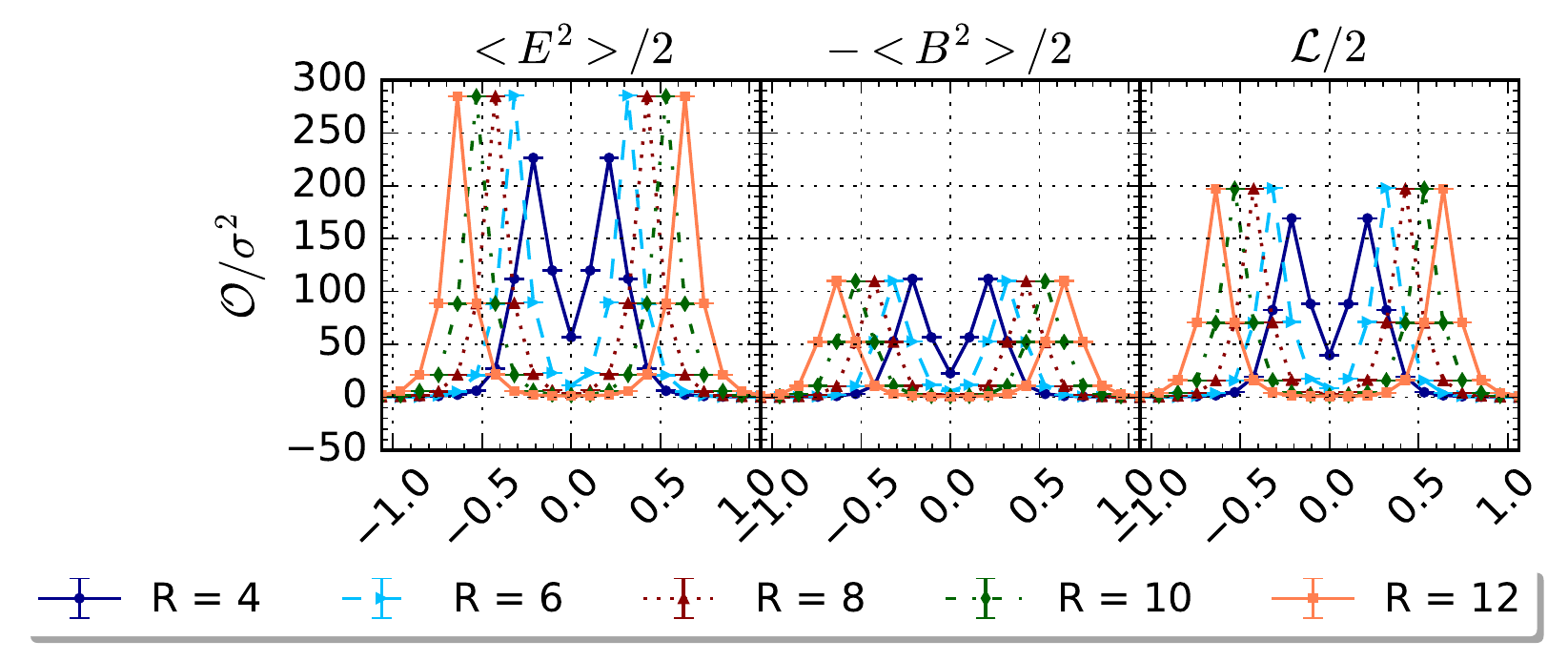}
\includegraphics[trim=56 30 10 0, clip,width=6.2cm]{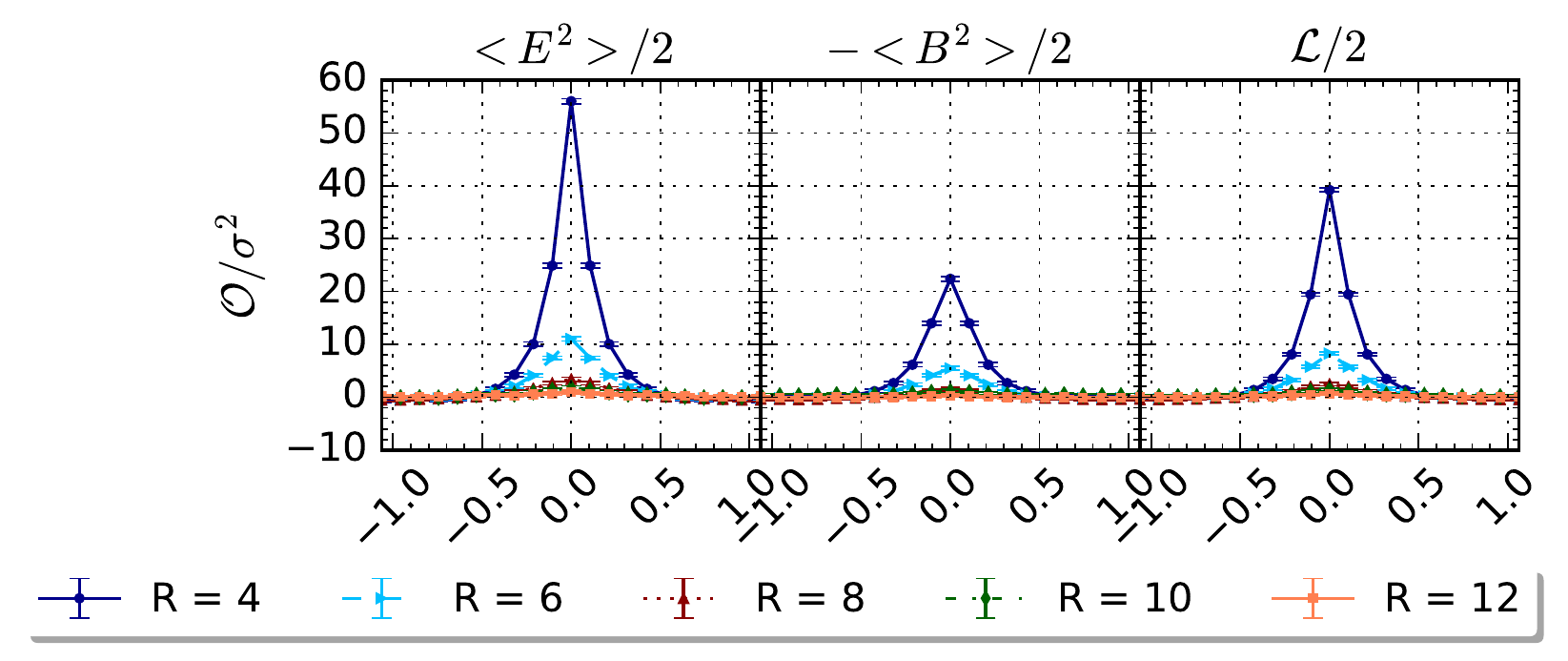}
\par\end{centering}}

\includegraphics[trim=5 0 10 170, clip,width=9.2cm]{F00_6.5_XZ_ppdagger_}
\par\end{centering}
\caption{The results for the $Q\bar{Q}$ system. The results in the left column correspond to the fields along the sources (plane XY) and the right column to the results in the middle of the flux tube (plane XZ). $R$ is the distance between the sources in lattice units.}
\label{fig:shapeFluxTube_QQbar}
\end{figure}

\begin{figure}[htp]
\captionsetup[subfloat]{farskip=0.1pt,captionskip=0.1pt}
\begin{centering}
\includegraphics[width=6.2cm]{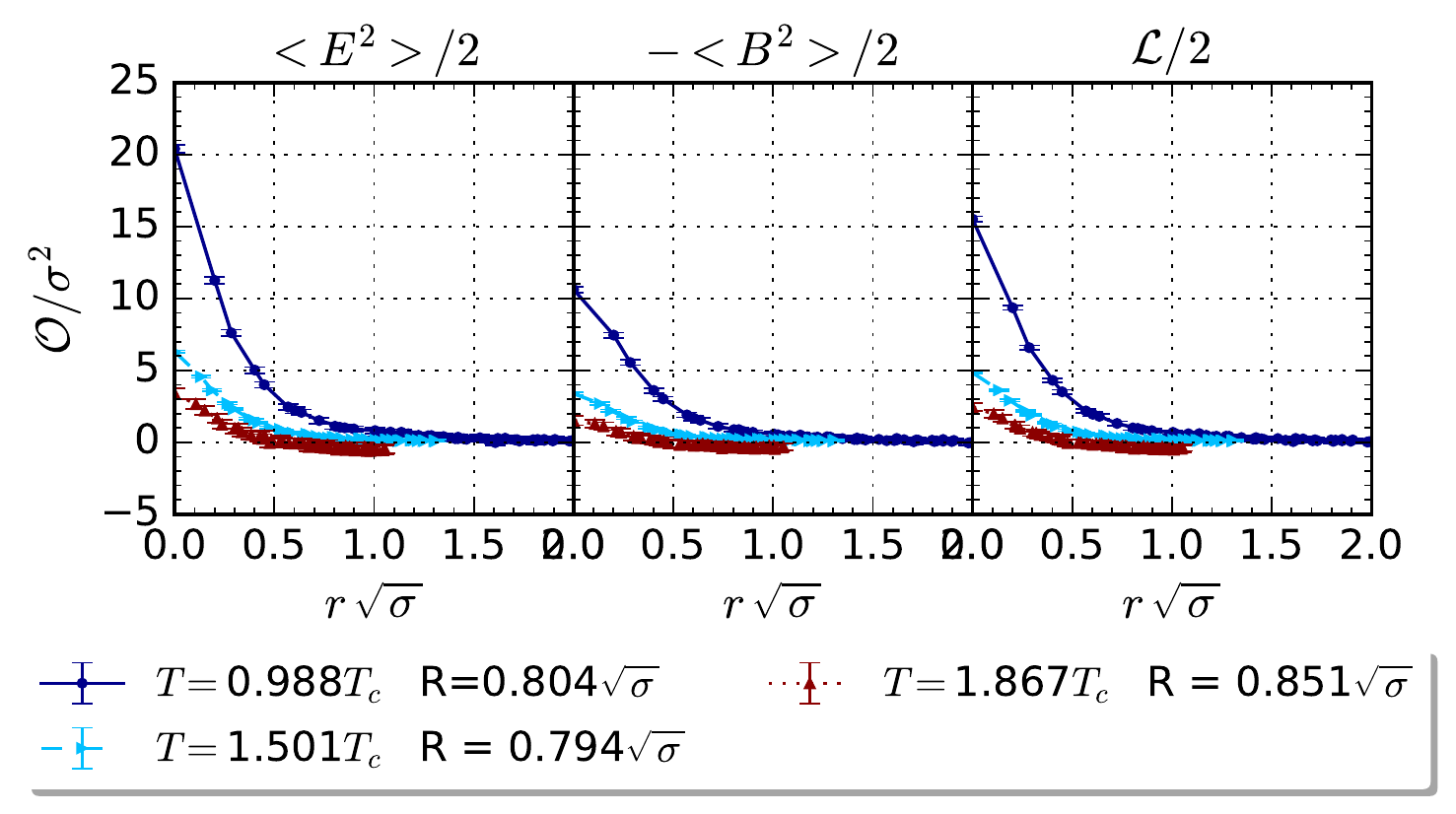}
\includegraphics[width=6.2cm]{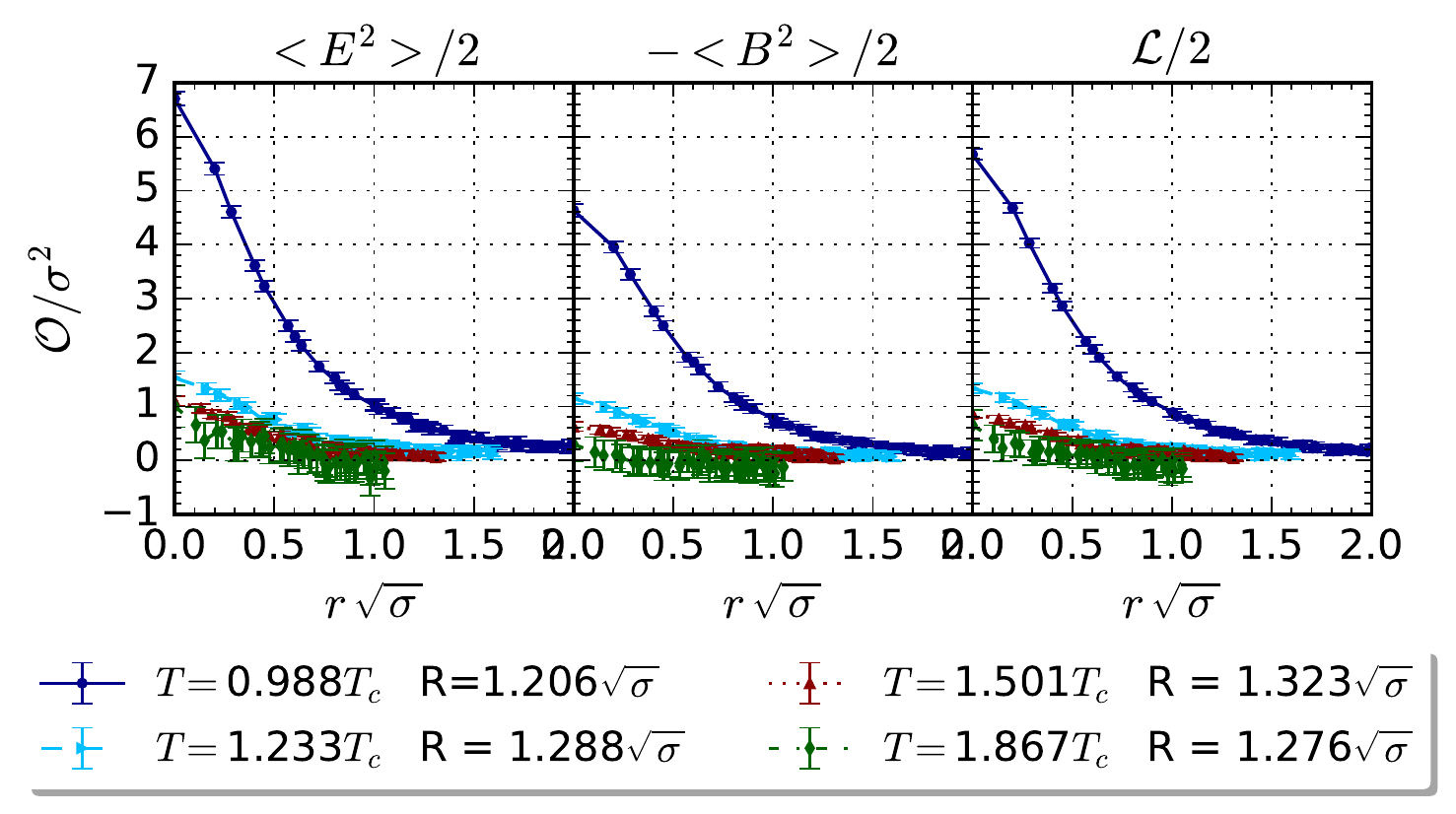}
\par\end{centering}
\caption{Results for the fields of the $Q\bar{Q}$ system in the middle of the flux tube in the plane XZ.}
\label{fig:shapeMidFluxTube_QQbar}
\end{figure}

The $QQ$ and $Q\bar{Q}$ are located at $(0,-R/2,0)$ and $(0,R/2,0)$ for $R=4,6,8,10\text{ and } 12$ lattice spacing units.
In \cref{fig:shapeFluxTube_QQbar,fig:shapeMidFluxTube_QQbar}, we show the results for the $Q\bar{Q}$ system. As expected the strength of the fields decrease with the temperature. Also, in the confined phase the width in the middle of the flux tube increases with the distance between the sources, while above the phase transition the width decreases with the distance.

\begin{figure}[htp]
\centering
\includegraphics[width=8cm]{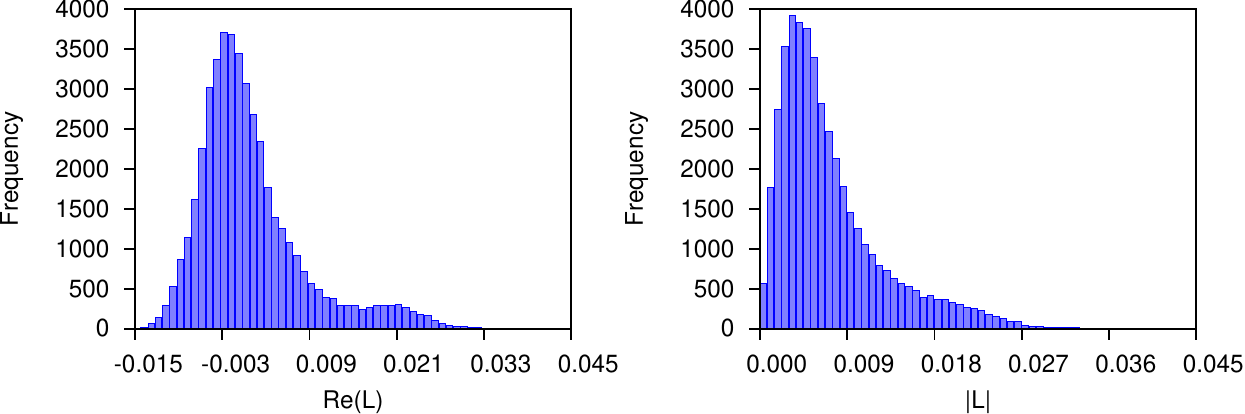}
\caption{Histogram of the Polyakov loop history for $\beta=6.055$.}
\label{histpolyloop}
\end{figure}

\begin{figure}[htp]
	\captionsetup[subfloat]{farskip=0.5pt,captionskip=0.5pt}
\begin{centering}

\subfloat[$\beta=6.055$, $T=0.988T_c$, with contaminated configurations.\label{fig:F00_6.055_pp_}]{
\begin{centering}
\includegraphics[trim=56 30 10 0, clip,width=6.2cm]{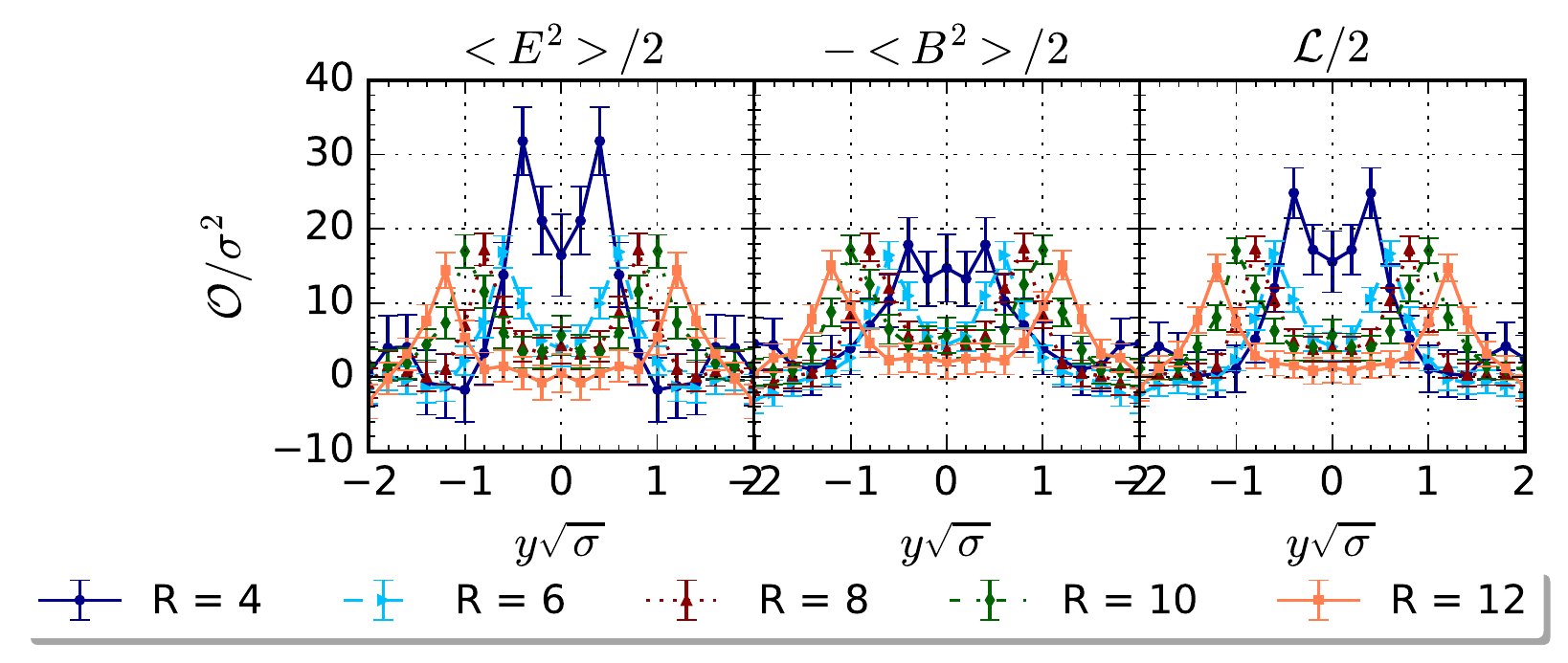}
\includegraphics[trim=56 30 10 0, clip,width=6.2cm]{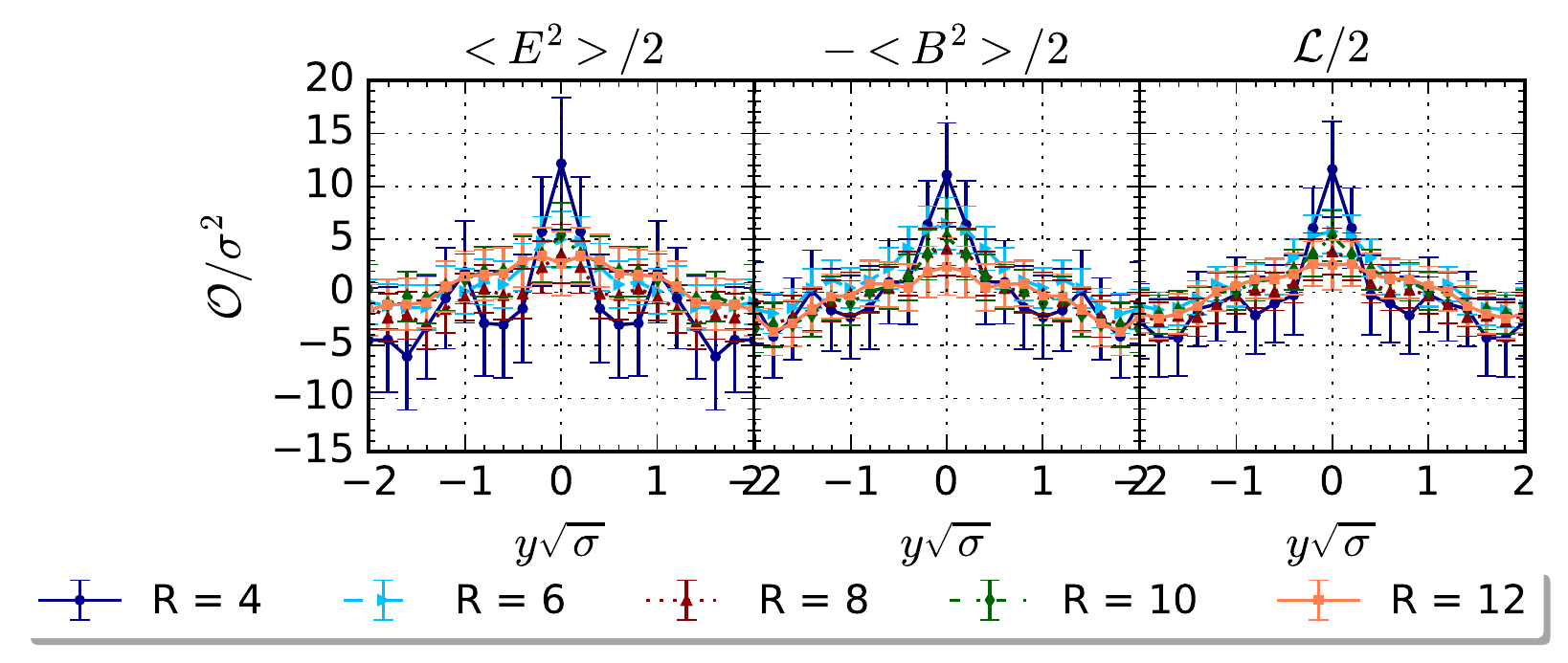}
\par\end{centering}}

\subfloat[$\beta=6.055$, $T=0.988T_c$, without contaminated configurations.\label{fig:F00_clean_6.055_pp_}]{
\begin{centering}
\includegraphics[trim=56 30 10 0, clip,width=6.2cm]{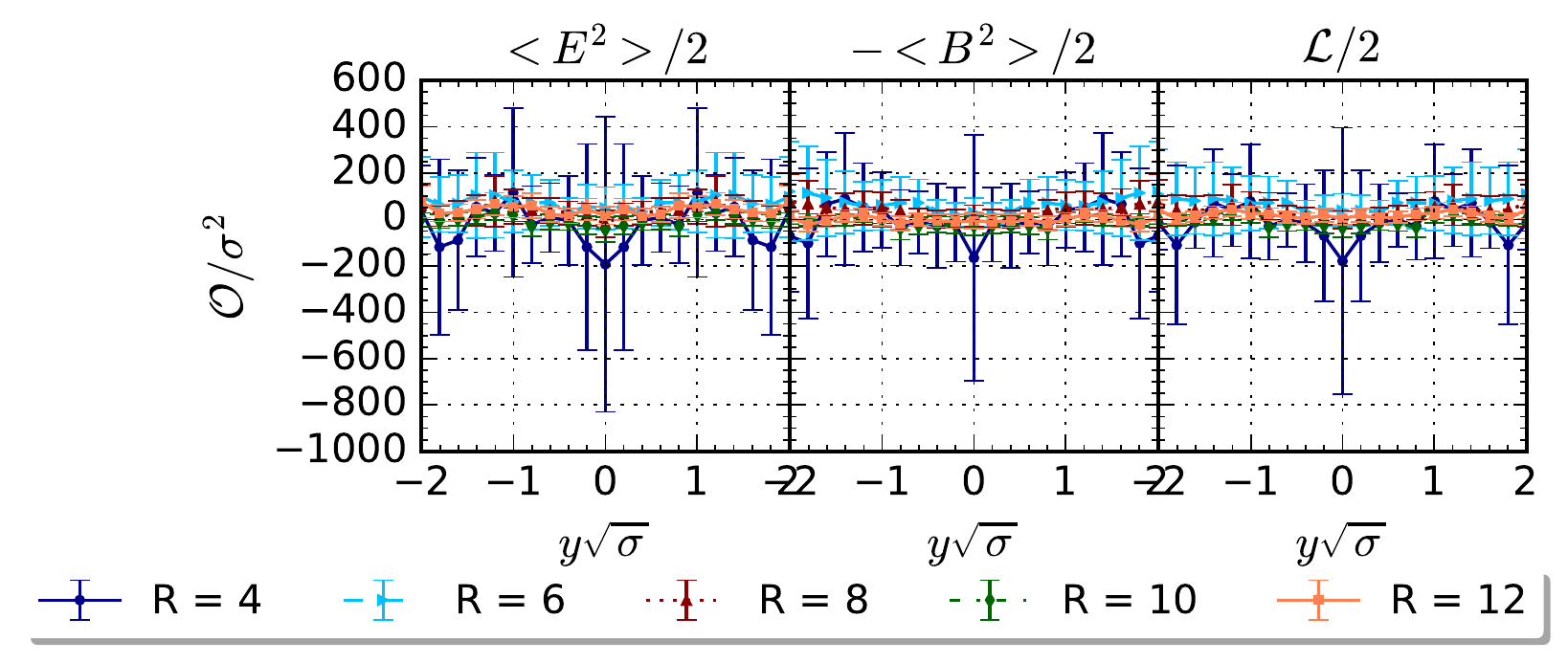}
\includegraphics[trim=56 30 10 0, clip,width=6.2cm]{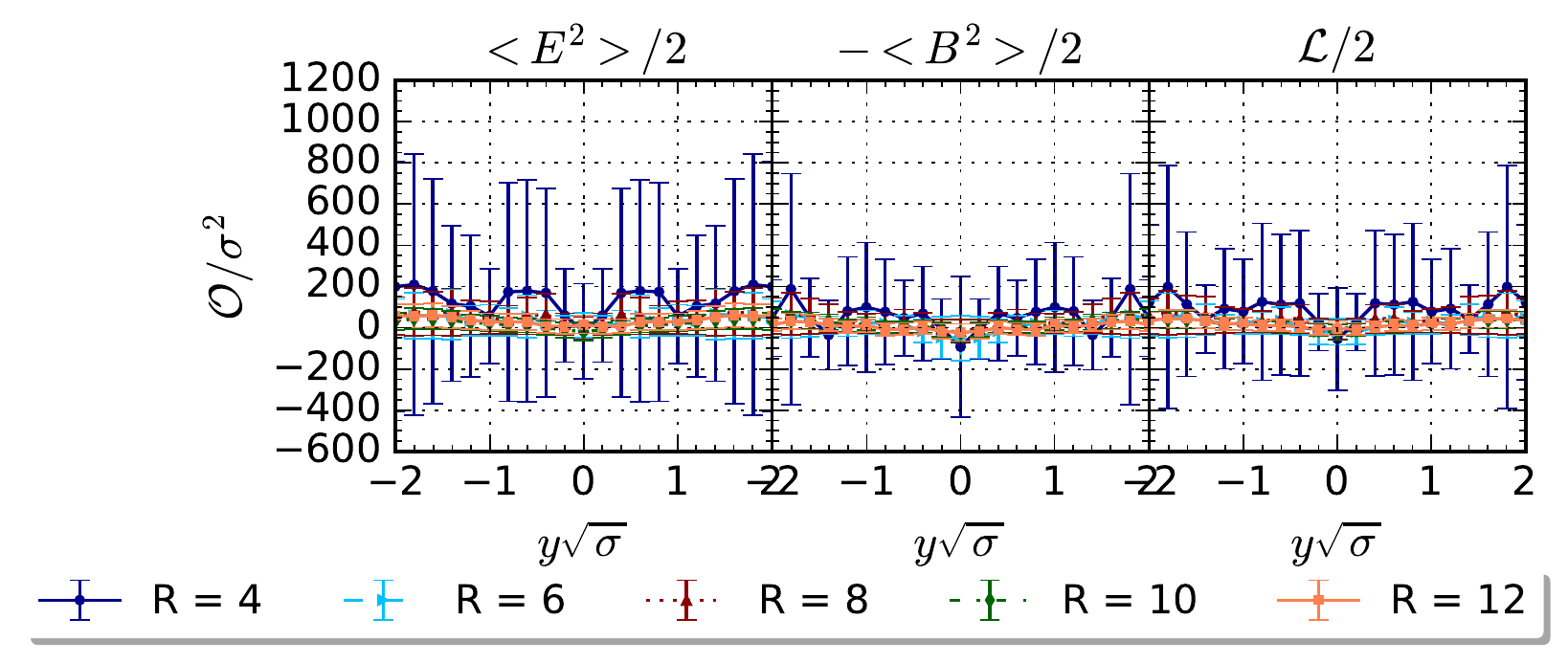}
\par\end{centering}}

\subfloat[$\beta=6.2$, $T=1.233T_c$.\label{fig:F00_6.2_pp_}]{
\begin{centering}
\includegraphics[trim=56 30 10 0, clip,width=6.2cm]{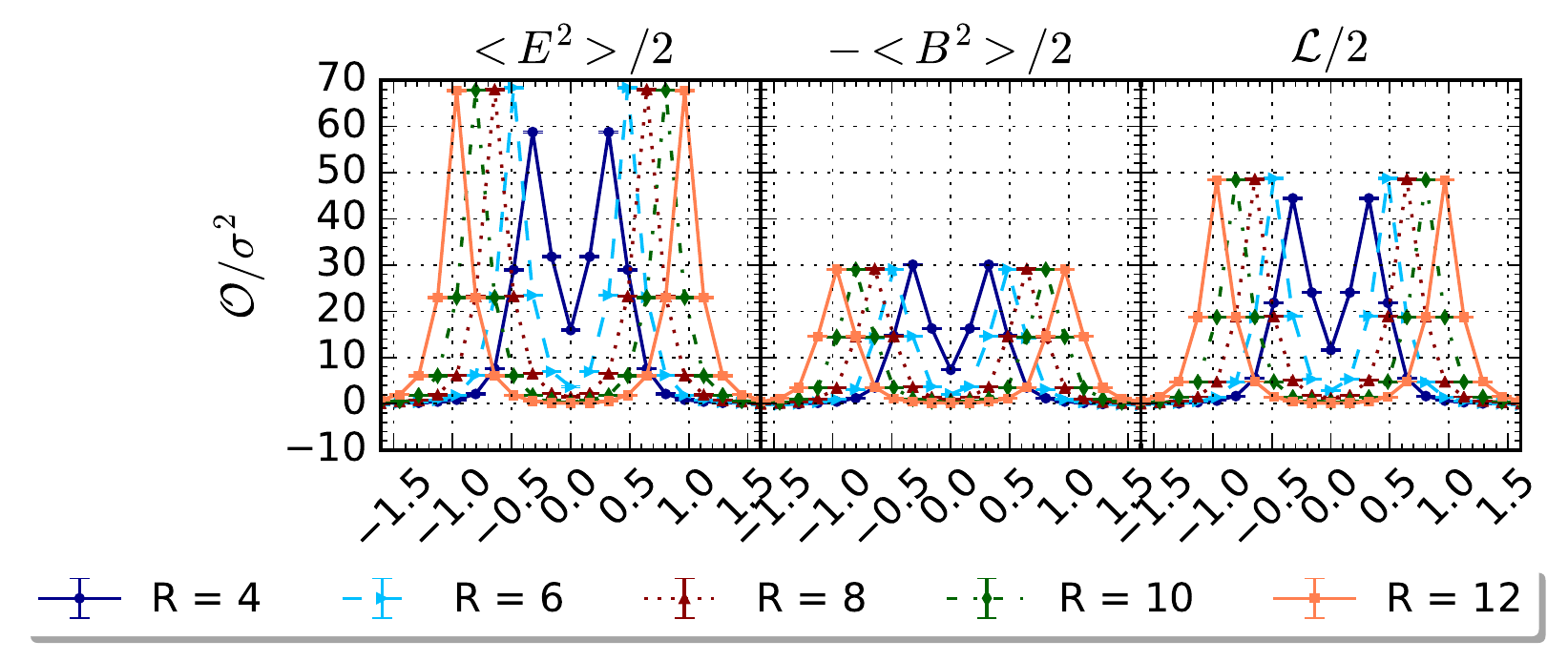}
\includegraphics[trim=56 30 10 0, clip,width=6.2cm]{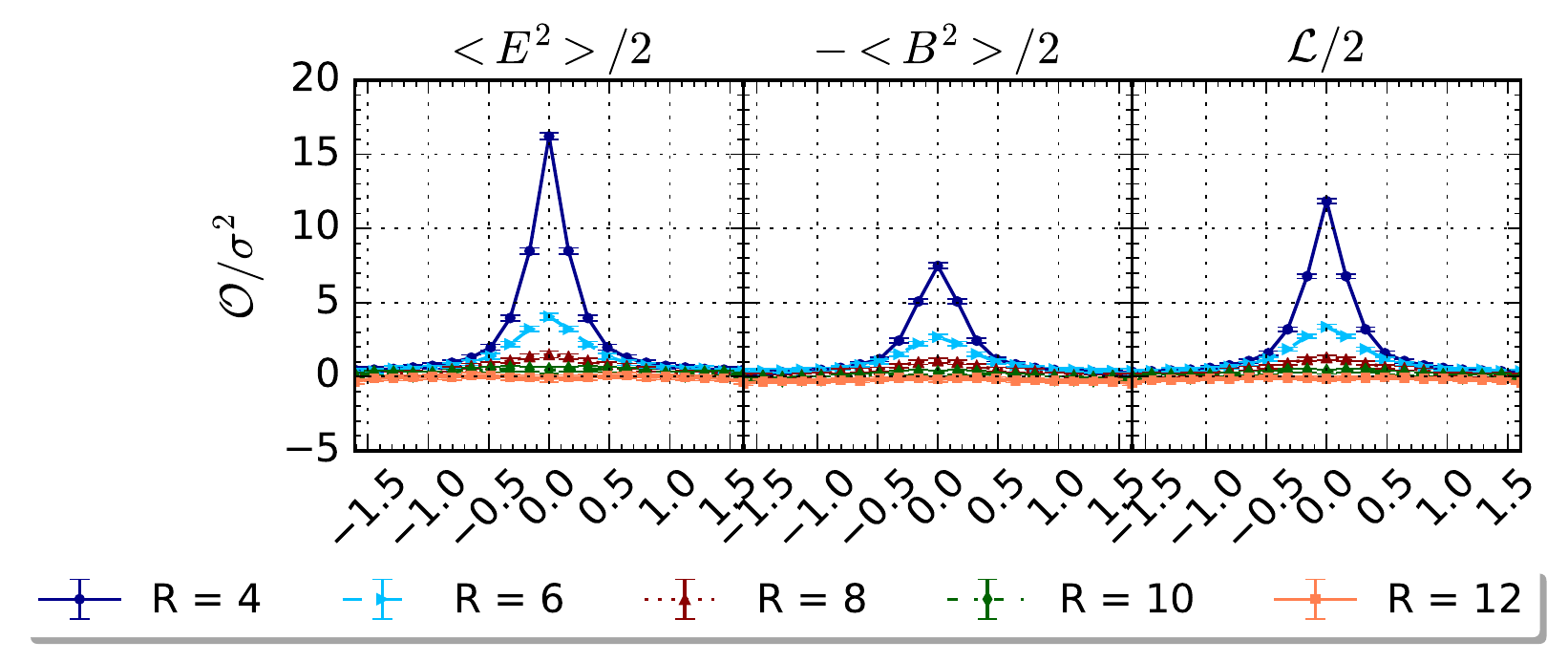}
\par\end{centering}}

\subfloat[$\beta=6.5$, $T=1.868T_c$.\label{fig:F00_6.5_pp_}]{
\begin{centering}
\includegraphics[trim=56 30 10 0, clip,width=6.2cm]{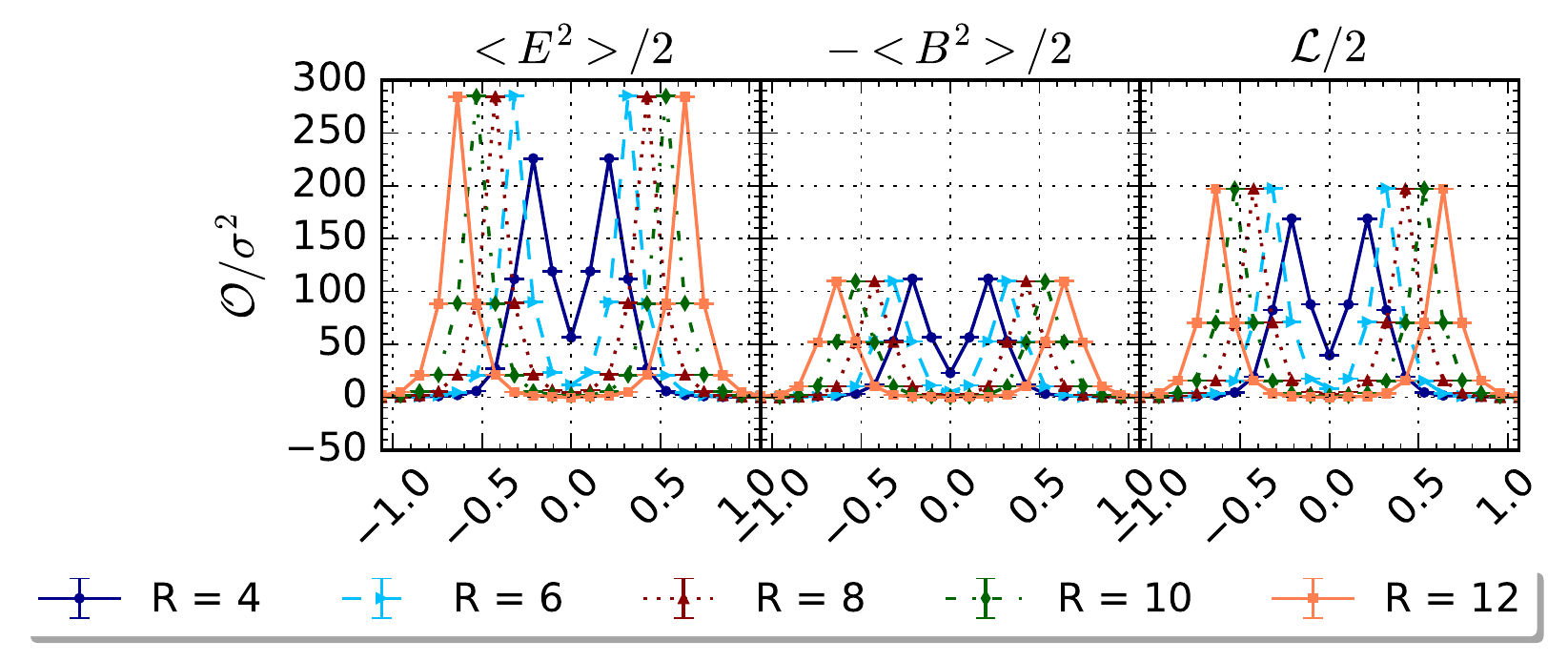}
\includegraphics[trim=56 30 10 0, clip,width=6.2cm]{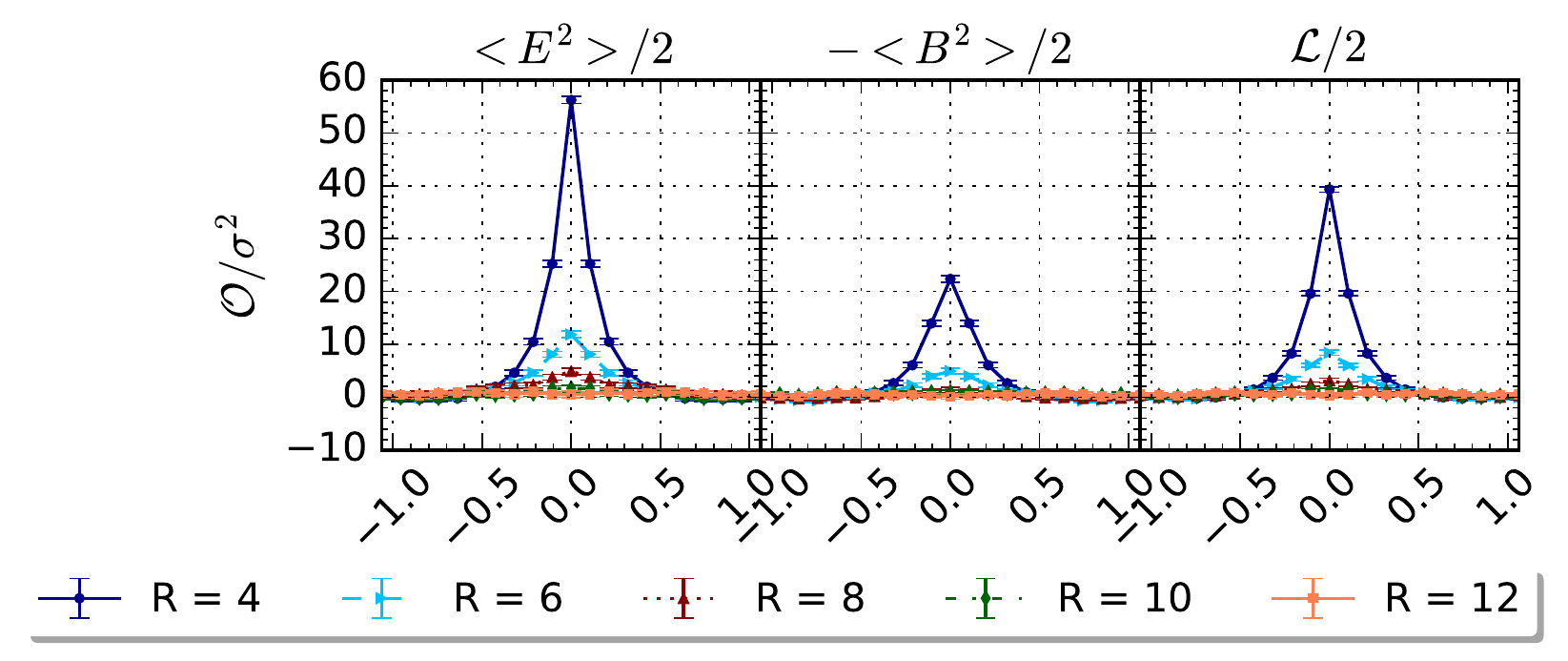}
\par\end{centering}}

\includegraphics[trim=5 0 10 170, clip,width=9.2cm]{F00_6.5_XZ_pp_}
\par\end{centering}
\caption{The results for the $QQ$ system. The results in the left column correspond to the fields along the sources (plane XY) and the right column to the results in the middle of the flux tube (plane XZ). $R$ is the distance between the sources in lattice units.}
\label{fig:shapeFluxTube_QQ}
\end{figure}

Just below the phase transition, we need to make sure that we don't have contaminated configurations as already mentioned in \cite{Cardoso:2011hh}. By plotting the histogram of  Polyakov loop history for $\beta=6.055$, \cref{histpolyloop}, we were able to identify a second peak which then we were able to remove all the configurations that lie on the second peak. Therefore, in \cref{tab:latticesimdata} the value with asterisk corresponds to the configurations after removing these contaminated configurations. In \cref{fig:F00_6.055_pp_,fig:F00_clean_6.055_pp_}, we show the results of this effect for the  $QQ$ system below the phase transition.

\begin{figure}[!htp]
	\captionsetup[subfloat]{farskip=0.1pt,captionskip=0.1pt}
	\begin{centering}
		\subfloat[$Q\bar{Q}$.\label{fig:Fr_ppdagger}]{
			\begin{centering}
				\includegraphics[width=6.2cm]{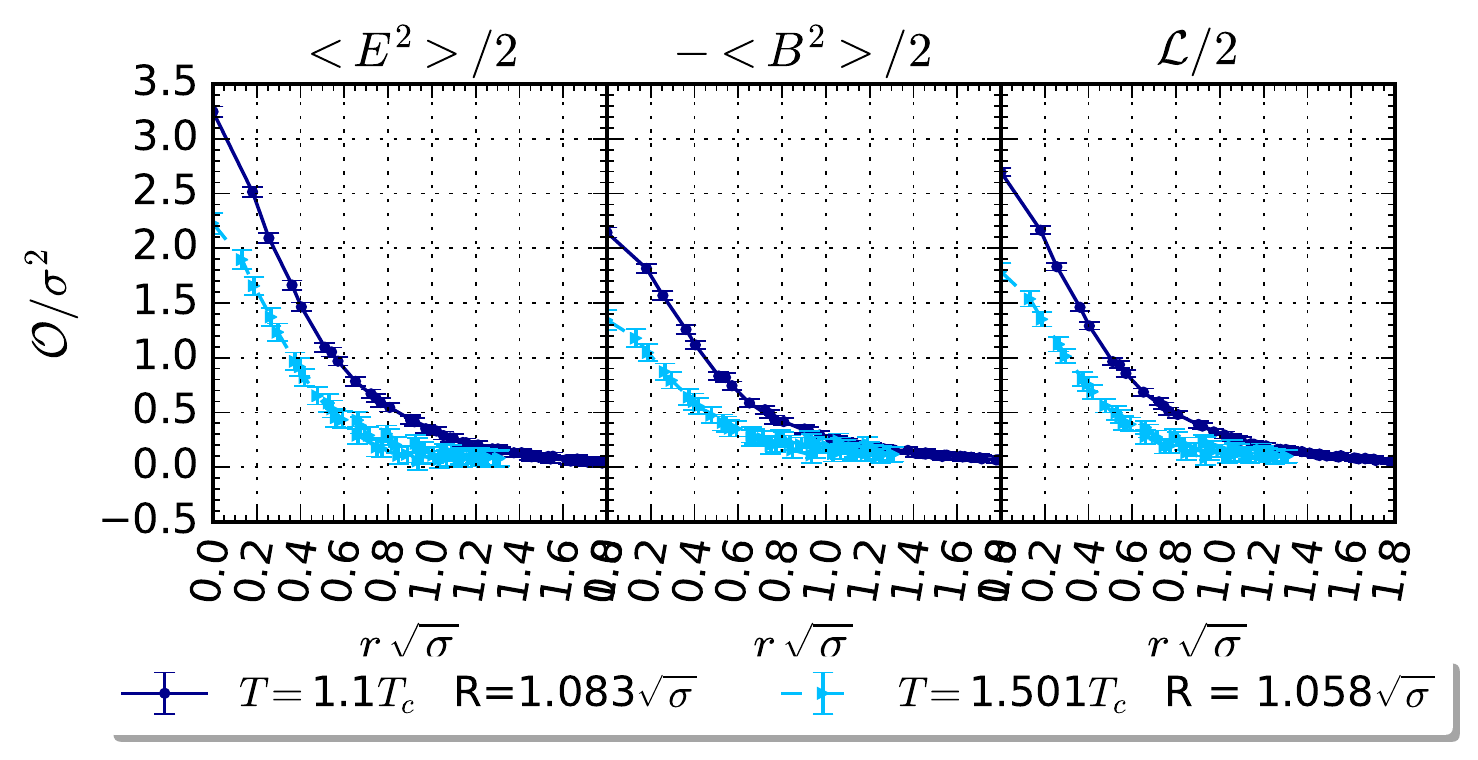}
				\par\end{centering}}
		\subfloat[$QQ$.\label{fig:Fr_pp}]{
			\begin{centering}
				\includegraphics[width=6.2cm]{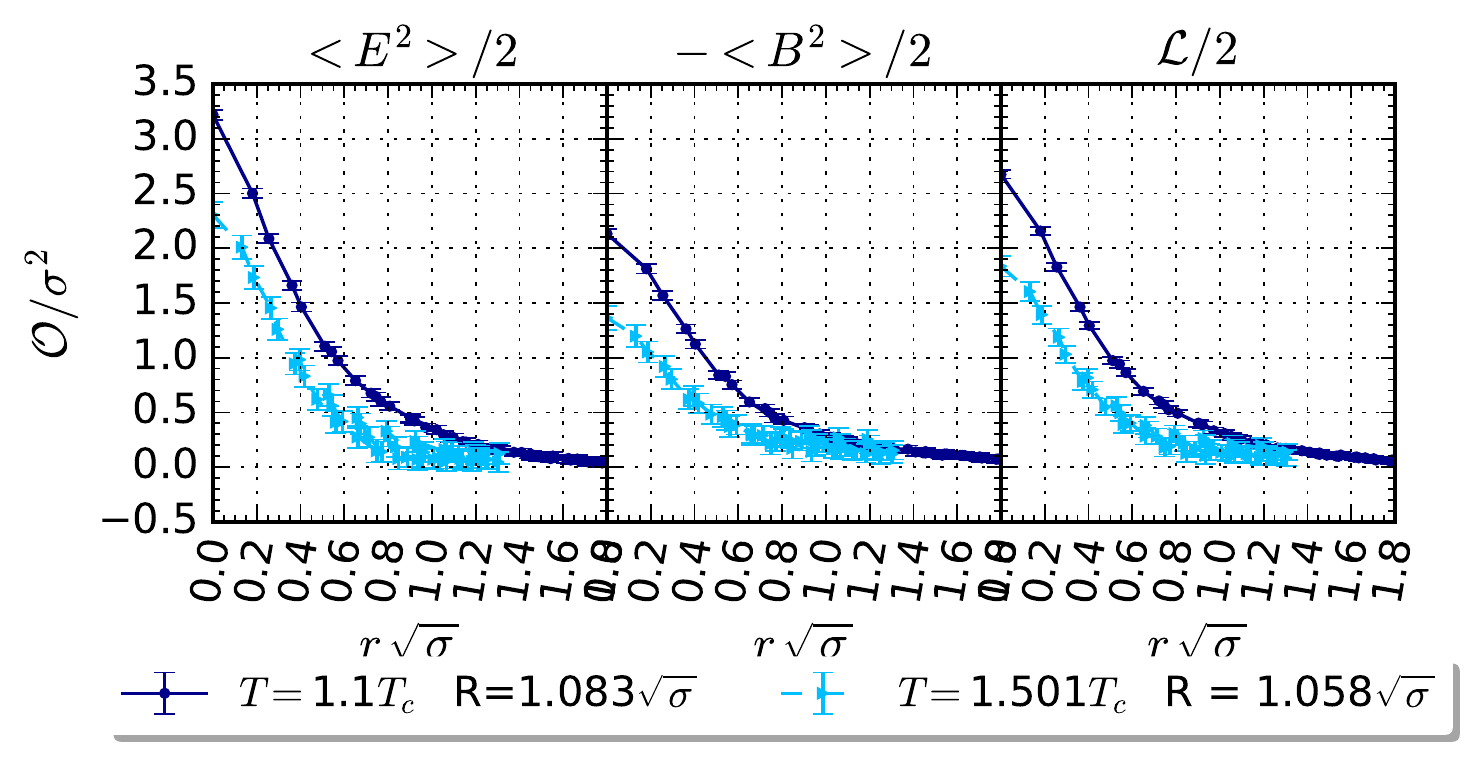}
				\par\end{centering}}
		\par\end{centering}
	\caption{Results for the fields in the middle of the flux tube in the plane XZ.}
	\label{fig:shapeMidFluxTube_QQ_QQbar}
\end{figure}

\section{Conclusions}

As the distance increase between the sources, the field strength at the flux tube decreases as already seen in studies at zero temperature.
Below the phase transition, the fields  strength decreases as the temperature increases. However, above the phase transition the fields rapidly decrease to zero as the quarks are pulled apart.
The width of the flux tube below the phase transition increases with the separation between the quark-antiquark, however above the phase transition the width seems to decrease.

\section*{Acknowledgments}


Nuno Cardoso and Marco Cardoso are supported by FCT under the contracts SFRH/BPD/109443/2015 and SFRH/BPD/73140/2010 respectively.
We also acknowledge the use of CPU and GPU servers of PtQCD, supported by NVIDIA, CFTP and FCT grant UID/FIS/00777/2013.


\end{document}